\documentclass[journal]{vgtc}
\onlineid{0}
\vgtccategory{Research}
\vgtcpapertype{please specify}

\graphicspath{{fig/}{figs/}{figures/}{pictures/}{images/}{./}}
\usepackage{tabu}                      
\usepackage{booktabs}                  
\usepackage{lipsum}                    
\usepackage{mwe}                       
\usepackage{mathptmx}                  

\title{DTBIA: An Immersive Visual Analytics System for Brain-Inspired Research}

\author{
      Jun-Hsiang\ Yao,
      Mingzheng Li,
      Jiayi Liu, 
      Yuxiao Li,
      Jielin Feng,
      Jun Han,
      Qibao Zheng,
      Jianfeng Feng, and 
      Siming Chen
}

\authorfooter{
    \item
    Jun-Hsiang Yao, Minzheng Li, Jiayi Liu, Jielin Feng, and Siming Chen are with School of Data Science, Fudan University. E-mail: rxyao24@m.fudan.edu.cn, simingchen@fudan.edu.cn. Siming Chen is the corresponding author.
    \item 
    Qibao Zheng and Jianfeng Feng are with Science and Technology for Brain-inspired Intelligence, Fudan University. E-mail: {zhengqb, jffeng}@fudan.edu.cn
    \item
    Yuxiao Li is with The Ohio State University. E-mail: li.14025@osu.edu.
    \item
    J. Han is with The Hong Kong University of Science and Technology and CORE. E-mail: hanjun@ust.hk.

}

\abstract{
The Digital Twin Brain (DTB) is an advanced artificial intelligence framework that integrates spiking neurons to simulate complex cognitive functions and collaborative behaviors. For domain experts, visualizing the DTB's simulation outcomes is essential to understanding complex cognitive activities.
However, this task poses significant challenges due to DTB data's inherent characteristics, including its high-dimensionality, temporal dynamics, and spatial complexity.
To address these challenges, we developed DTBIA, an Immersive Visual Analytics System for Brain-Inspired Research.
In collaboration with domain experts, we identified key requirements for effectively visualizing spatiotemporal and topological patterns at multiple levels of detail.
DTBIA incorporates a hierarchical workflow — ranging from brain regions to voxels and slice sections — along with immersive navigation and a 3D edge bundling algorithm to enhance clarity and provide deeper insights into both functional (BOLD) and structural (DTI) brain data.
The utility and effectiveness of DTBIA are validated through two case studies involving with brain research experts. The results underscore the system's role in enhancing the comprehension of complex neural behaviors and interactions.

}


\teaser{
  \centering
  \includegraphics[width=\linewidth]{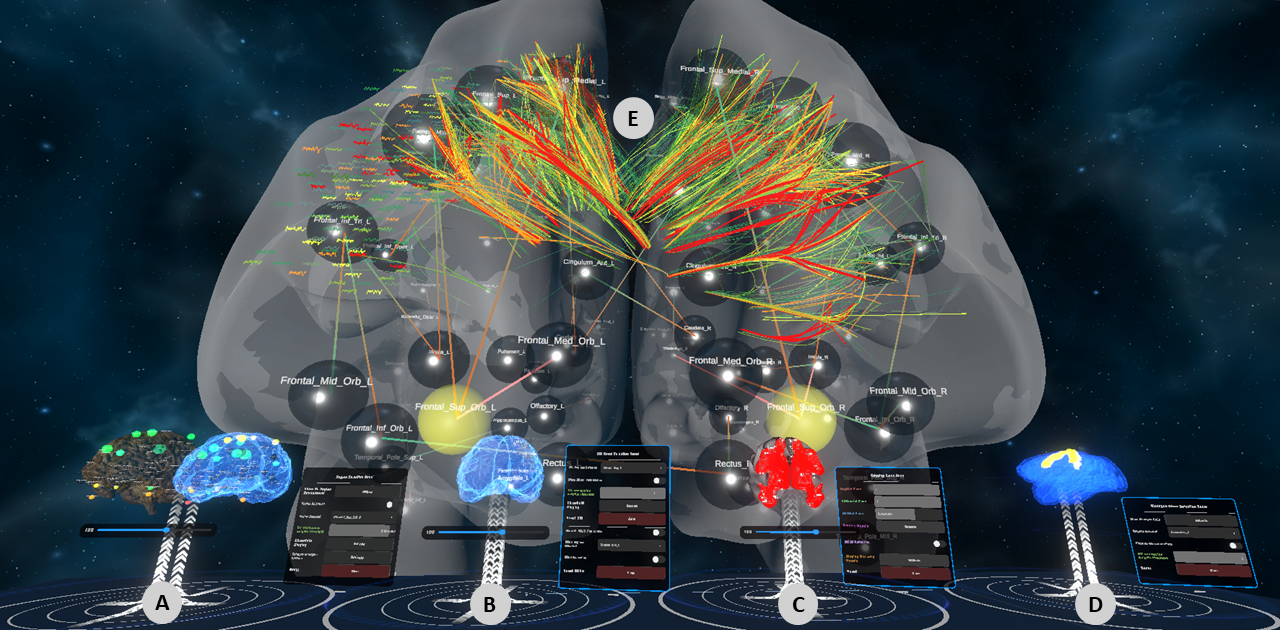}
  \caption{%
    The DTBIA system offers a multi-level exploration framework, incorporating three hierarchical processes: (A) Region-Level, (B) Voxel-Level, and (C) Slice-Section-Level for human brain data, along with a model (D) for macaque brain data. For immersive exploration, DTBIA utilizes two distinct brain models: a Real-Scale brain, positioned near the edge of the screen for functional data analysis, and a Large-Scale brain (E) for structural data exploration, enhanced by the FDEB algorithm to improve clarity and connectivity analysis in the navigation space.
  }
  \label{fig:teaser}
}

\begin{document}
\firstsection{Introduction}
\maketitle

Brain-inspired computing models\cite{zhang2020system, parhi2020brain} are a form of artificial intelligence (AI)\cite{savage2019ai} that integrate hardware and software to simulate the information processing and learning mechanisms of the human brain\cite{jain2023ai}. 
These models aim to develop more intelligent and efficient AI systems by constructing neural-based computational units and establishing their interconnections\cite{li2024brain}. 
However, current brain-inspired models face limitations due to incomplete understanding of brain functions and insufficient computational methods for processing various neural signals\cite{feng2024human}. 
This often results in potential inaccuracies in simulations, alongside the requirement for significant computational resources and large datasets for training and optimization.

To address these issues, Lu et al. \cite{DTB} and their team, the experts who collaborated with us, have proposed a new framework called the Digital Twin Brain (DTB), designed to meticulously emulate the intricate functions and behaviors of neurons in the human brain\cite{lu2023digital}.
The DTB acts as a bridge between neuroscience and AI by integrating biological brain networks with computational models, utilizing biological insights to enhance artificial systems while employing AI to advance the understanding of biological mechanisms, fostering a bidirectional synergy between the two domains\cite{xiong2023digital}.
This innovative framework goes beyond basic simulation to calculate functional neuroanatomy, focusing on Blood-Oxygen-Level-Dependent (BOLD) signals at the voxel level. 
The dataset comprises over 22,703 time series, each spanning 800 ms and providing critical volumetric observations for understanding neural activity. Furthermore, the DTB data reveals an extensive network of neural pathways, elucidating the topological connections among brain regions and voxels, which are crucial for comprehending the brain's intricate architecture.

Despite the richness of DTB data, traditional visualization and analysis methods struggle to accommodate its complexity and volume\cite{li2023dtbvis}. 
Over the past one and a half years, we have engaged deeply with domain experts from the Institute of Science and Technology for Brain (ISTB) at Fudan University, collaborating to address their DTB visualization needs. 
We organized their requirements to compare the simulation results from the DTB with biological brain data to evaluate the models' effectiveness while analyzing functional and structural data patterns in both biological and digital brains. 
In this process, we identified that existing visualization tools face three main challenges: Inefficient Data Exploration, Inadequate Spatial Awareness, and Overwhelming Visual Complexity. These challenges hinder effective analysis of dense, multidimensional brain data, resulting in reduced expert engagement and insight.

To address these challenges, we introduce DTBIA, an immersive visual analytics system designed to enhance clarity and usability in the processing, analysis, and visualization of large-scale, high-dimensional DTB data. It achieves this through hierarchical, top-down exploration within an immersive environment, combined with advanced 3D edge bundling techniques to optimize data representation. 
DTBIA adopts a hierarchical, top-down exploration strategy, allowing users to transition seamlessly from macro-level overviews to detailed analyses of specific brain regions and voxels. The system integrates multiple data levels, including the whole brain, regions, and individual voxels, and provides interactive tools that allow users to specify brain regions, time points, and sectional slices for focused analysis. 
In addition, DTBIA leverages Immersive Virtual Environments (IVEs) to provide a Real-Scale brain model for broad overviews and a Large-Scale brain model for navigation and exploration. Unlike traditional 2D systems, Virtual Reality (VR) enables embodied interaction, allowing users to navigate the brain's 3D structure through egocentric walking for Real-Scale views and egocentric flying for detailed exploration. This immersive approach enhances spatial awareness and reveals spatiotemporal patterns and insights.
To further reduce visual clutter and improve the clarity of neural pathway representations, DTBIA incorporates 3D edge bundling techniques. These techniques cluster dense neural connections, minimize visual noise, and emphasize key pathways, particularly those derived from Diffusion Tensor Imaging (DTI) data. 
The effectiveness of the DTBIA system is validated through two real-world case studies, demonstrating its practical utility in advancing DTB research.

\par Our study provides the following research contributions:
\begin{itemize}
    \item We collaborated with 12 domain experts to address the gap in effective DTB visualization tools. This interdisciplinary effort bridges brain-inspired computing and neuroscience, facilitating better insight exchange and deepening the role of visualization in advancing brain-inspired research.
    
    \item We developed an immersive visual analytics system that leverages immersive visualization techniques to explore DTB data, facilitating the analysis of spatiotemporal patterns and data behavior.
    
    \item We have demonstrated the practical value of our system for DTB research through the validation and analysis of real-world case studies, as well as expert feedback and evaluation.
\end{itemize}
\section{Related work}
This section reviews the literature across three key domains: Brain-Inspired Data Visualization for connecting biological insights with computational models, Immersive Analytics for intuitive 3D exploration, and Spatiotemporal Visualization for analyzing patterns across time and space.

\subsection{Brain-Inspired Data Visualization}
The need to simulate the human brain has driven decades of interdisciplinary research at the intersection of neuroscience and AI~\cite{feng2024human}, laying the foundation for brain-inspired computing~\cite{parhi2020brain, zhang2020system, li2024brain}. While early AI was inspired by neural networks modeled on the brain, advances in neuroscience now enable biologically realistic models that can bridge the gap between biology and AI~\cite{jain2023ai, xiong2023digital}. Among these efforts, the Digital Twin Brain (DTB) framework~\cite{lu2023digital, DTB}, developed by ISTB at Fudan University in collaboration with our team, stands out as a comprehensive simulation and assimilation platform for modeling the whole human brain. Unlike other frameworks~\cite{markram2006blue, amunts2016human}, such as The Virtual Brain (TVB)~\cite{sanz2013virtual}, which focuses on mesoscale dynamics, or SpiNNaker~\cite{furber2014spinnaker, sen2018building}, which emphasizes spiking neural networks, DTB integrates multimodal data from structural MRI~\cite{esteban2019fmriprep}, DTI, and PET scans to capture heterogeneity and coupling at various scales. On the hardware side, DTB employs GPU-based high-performance computing clusters with advanced load balancing and data traffic optimization, enabling simulations of 86 billion neurons and 47.8 trillion synapses at unprecedented speed, while addressing the scalability and fidelity limitations of traditional systems~\cite{lu2023digital, DTB}.  This combination of computational power and biological realism positions DTB as a pioneering tool for advancing brain-inspired research~\cite{xiong2023digital}.

Visualization plays a vital role in neuroscience by enabling intuitive analysis of complex brain data. Foundational efforts like BrainMap~\cite{PMID:11967563} and the Human Connectome Project~\cite{HumanConnectome} provided large-scale neuroimaging databases and multimodal brain mapping. Tools such as BrainBrowser~\cite{Brainbrowser}, BrainGL~\cite{brainGL}, and BrainViewer~\cite{Brainviewer} allow interactive exploration of structural and functional data, while Brain3X~\cite{10.3389/fninf.2015.00002} enables real-time interaction with brain network dynamics. Specialized systems like NeuroLines~\cite{6875935} and NeuroTessMesh~\cite{10.3389/fninf.2017.00038} focus on neural connections and 3D morphology, and NIVR~\cite{7892381} extends brain visualization into virtual environments.

Despite the rich literature on brain visualization, few efforts have addressed the specific challenges of visualizing DTB data. To fill this gap, Li et al., one of our coauthors, developed DTBVis~\cite{li2023dtbvis}, a system designed to support DTB research with features tailored to hierarchical data representation and spatiotemporal analysis. While DTBVis lays an important foundation for comparing DTB datasets, our work introduces a significant advancement by integrating immersive environments and enhanced topology pattern analysis. This underscores the need for specialized visualization systems that meet the unique demands of DTB research, thereby advancing both neuroscience and brain-inspired computing.

\subsection{Immersive Analytics}
Immersive Analytics (IA)~\cite{marriott2018immersive, Dwyer2018} leverages advanced technologies such as Virtual Reality (VR) and Augmented Reality (AR) to enhance data exploration and analysis by integrating users into data-rich environments~\cite{zhao2022metaverse}. Unlike traditional analytics, IA capitalizes on human perceptual abilities, including spatial reasoning, proprioception, and kinesthetic awareness, to foster intuitive interaction and a deeper understanding of complex data sets~\cite{IAsurvey1, IAsurvey4}. This paradigm shift enables analysts to "step through the glass"~\cite{ens2021grand}, engaging directly with multidimensional data representations.

IA systems address diverse challenges across various fields. Zhang et al.\cite{9756734} introduced TimeTables, a VR-based platform for exploring spatiotemporal data, while Spur et al.\cite{MapStack} developed MapStack, enabling multi-level geospatial data comparison in VR. Cunningham et al.\cite{8533896} proposed an immersive law enforcement tool for crime data analysis, and Chu et al.\cite{9557225} created Tivee for analyzing badminton strategies using 3D interactive visualizations. Ye et al.\cite{9222313} extended this with ShuttleSpace, which maps badminton trajectories into 3D immersive environments. Zhang et al.\cite{ZHANG2021128} introduced UrbanVR, which visualizes large-scale urban data in VR as 3D scenes, providing an intuitive interface for interaction. Toolkits like IATK~\cite{BrainIA6} and DXR~\cite{BrainIA7} enable building customizable immersive data visualizations, highlighting the flexibility of IA across disciplines.

In neuroscience, IA addresses the challenges of visualizing complex brain data. For example, NeuroCave~\cite{10.1162/netn_a_00044} provides an immersive platform for exploring brain connectivity, enabling researchers to intuitively analyze neural structures. Hellum et al. introduced SONIA~\cite{BrainIA3}, a VR tool designed for brain network exploration, while De Ridder et al.~\cite{BrainIA4} applied immersive environments to visualize functional MRI (fMRI) data, and Shattuck et al.~\cite{shattuck2018multiuser} developed a multi-user VR system for collaborative neuroimaging analysis. These tools highlight IA's transformative potential in neuroscience.

While previous IA systems have focused on other application domains, our work directly addresses the unique challenges posed by brain-inspired research, advancing the field by enabling deep, immersive exploration of DTB data. This immersive approach offers a self-centered perspective and enhanced interactions, enabling users to better comprehend and analyze DTB data.

\subsection{Spatiotemporal Visualization}
Spatiotemporal visualization~\cite{zhu2021taxonomy} integrates temporal and spatial dimensions to uncover patterns, trends, and anomalies in complex data~\cite{chen2019survey}. Andrienko et al.\cite{Andrienko2003ExploratorySV} provided foundational methods, including map-based and multidimensional approaches, while the space-time cube\cite{geovisualization, Gatalsky2004InteractiveAO} remains a seminal framework for encoding spatial, temporal, and attribute data within a unified 3D framework. Thakur et al.\cite{3DMaps} expanded on this concept by integrating multiple time-series datasets into geographic space, facilitating the analysis of spatial-temporal relationships. Tominski et al.\cite{GW} proposed a 2D mapping approach, projecting spatiotemporal data along temporal and spatial axes for simplified exploration.

Applications span various domains. For urban analytics, Mota et al.\cite{mota2022comparison} compared 3D visualization methods for urban data, and Ferreira et al.\cite{6634127} developed a tool for taxi trip data analysis. In trajectory visualization, He et al.\cite{he2019variable} proposed a variable-based approach for spatiotemporal trajectories, while Nagel et al.\cite{8986941} introduced cpmViz for handling uncertainty in climate data. Tools like Voila by Cao et al.\cite{8022952} and DDLVis by Li et al.\cite{9552191} focus on real-time anomaly detection and density analysis. In neuroscience, Kasabov et al.\cite{kasabov2016mapping} leveraged spiking neural networks in the NeuCube framework to map and classify fMRI data, and Purgato et al.\cite{purgato2017interactive} developed interactive techniques for visualizing brain spatiotemporal networks. These methods demonstrate the critical role of tailored spatiotemporal solutions for understanding dynamic neural patterns.

While existing spatiotemporal visualization techniques tackle challenges across various domains, they do not address the hierarchical and dynamic complexities of DTB data. Our work contributes a novel immersive, multi-level spatiotemporal visualization method specifically designed to handle the unique needs of brain-inspired research, offering a more effective way to explore complex, multi-scale neural data.
\section{Overview}
In this section, we present an overview of the data used in our research, including hierarchical brain structures, functional data from BOLD signals, and structural data from DTI. These datasets form the foundation of our collaborative work with domain experts, which aimed to identify the challenges and needs of experts in analyzing high-dimensional brain data. The insights gained from this collaboration have guided the development of visualization tools tailored to the needs of DTB research.

\subsection{Data Description}
This subsection provides an overview of the hierarchical brain structure, functional data (BOLD signal), and structural data (DTI) used in our research. Both human and macaque brain datasets were utilized to support our analysis.

\subsubsection{Hierarchical Brain Structure}
The brain is divided into distinct regions, each associated with specific functions. For human brain studies, we rely on the Automated Anatomical Labeling (AAL) atlas \cite{TZOURIOMAZOYER2002273}. In AAL, the human brain is divided into 116 brain regions based on its anatomical features, with each region assigned a unique label and specific name for easy identification. These regions cover both hemispheres and include areas in the cerebral cortex, basal ganglia, limbic system, thalamus, brainstem, and cerebellum. However, since our focus is primarily on brain-related functions, we concentrate on 92 regions of the brain, excluding those associated with the cerebellum. Each brain region contains a varying number of voxels, which are the smallest three-dimensional units used to analyze brain data. Each voxel represents thousands of neurons, and Voxel-Level data enables detailed investigations of both functional and structural variations across different brain regions.

\subsubsection{BOLD Signal (Functional Data)}
The Blood-Oxygen-Level-Dependent (BOLD) signal, obtained through functional magnetic resonance imaging (fMRI), measures brain activity by detecting changes in blood oxygenation. The BOLD signal reflects the metabolic activity in the brain, as activated neurons consume more oxygen, leading to localized decreases in blood oxygenation that can be captured by fMRI scans. In our study, we used resting-state BOLD data from human participants and simulated data from DTB models. Resting-state refers to fMRI scans taken while participants are at rest, commonly used to study brain connectivity and organization~\cite{lv2018resting}. The dataset includes time-series data from 22,703 voxels, sampled every 800 milliseconds over 166 time points, enabling a detailed comparison of real and simulated BOLD signals to assess brain functionality."

\subsubsection{DTI (Structural Data)}
Diffusion Tensor Imaging (DTI) is a technique used to map the structural connectivity of the brain by measuring the diffusion patterns of water molecules along white matter tracts. The structural data in this study consists of voxel-wise connectivity matrices derived from DTI scans of both human and macaque brains. The dataset is represented by a 22,703 x 22,703 fully connected matrix, where each element reflects the probability of a connection between two voxel pairs. These connection probabilities, calculated through global normalization, provide a detailed measure of the strength of the structural connections. This DTI data enables us to explore the intricate network of anatomical connections between brain regions and their influence on brain function.


\subsection{Collaborative Feedback and Design Insights} 
This subsection outlines the iterative collaboration with domain experts to address the specific challenges and visualization needs in DTB research. Over the course of a year and a half, we worked closely with these experts, focusing on the specific visualization needs that emerge in their work. Throughout this period, we continuously refined and adapted our research tasks in response to their ongoing feedback, ensuring the system effectively met their evolving requirements. These interactions were critical in shaping both the design and functionality of the system, specifically addressing the complexities associated with analyzing high-dimensional brain data.

\textbf{Participants.} We engaged with 10 experts (E1-E10) from the Institute of Science and Technology for Brain-Inspired Intelligence at Fudan University. Among them, 8 specialize in developing mathematical and computational models that simulate brain functions, while 2 have a strong background in neuroscience. The regular meetings, held once or twice a week over a period of one and a half years, provided continuous insights that helped shape the development of the system. These meetings also enabled the experts to offer feedback on the system design, following a think-aloud protocol for real-time adjustments and enhancements.


During our collaborative discussions, experts provided continuous feedback on the challenges they encountered with existing visualization tools. For instance, E1 highlighted that while correlation metrics are typically used to assess model accuracy, there is a lack of interactive tools to comprehensively compare simulated neural signals with real recorded data. E3 emphasized the challenges in seamlessly tracking data across multiple modalities, further complicating analysis. These insights revealed key issues: First, as noted by E2, current tools are inefficient for exploring large-scale neural datasets, making it difficult to compare simulated and real-world signals. Additionally, several experts, including E4 and E7, stressed that 2D visualization methods lacked the spatial depth necessary to understand the complex relationships between brain regions, limiting effective data exploration. Furthermore, the overwhelming visual complexity, particularly in managing dense neural connections, emerged as a major challenge. As E5 observed, tools like Matplotlib struggle with high-dimensional functional and structural data, often leading to visual clutter and obscuring critical insights.

\begin{figure*}[htb]
     \includegraphics[width=\linewidth]{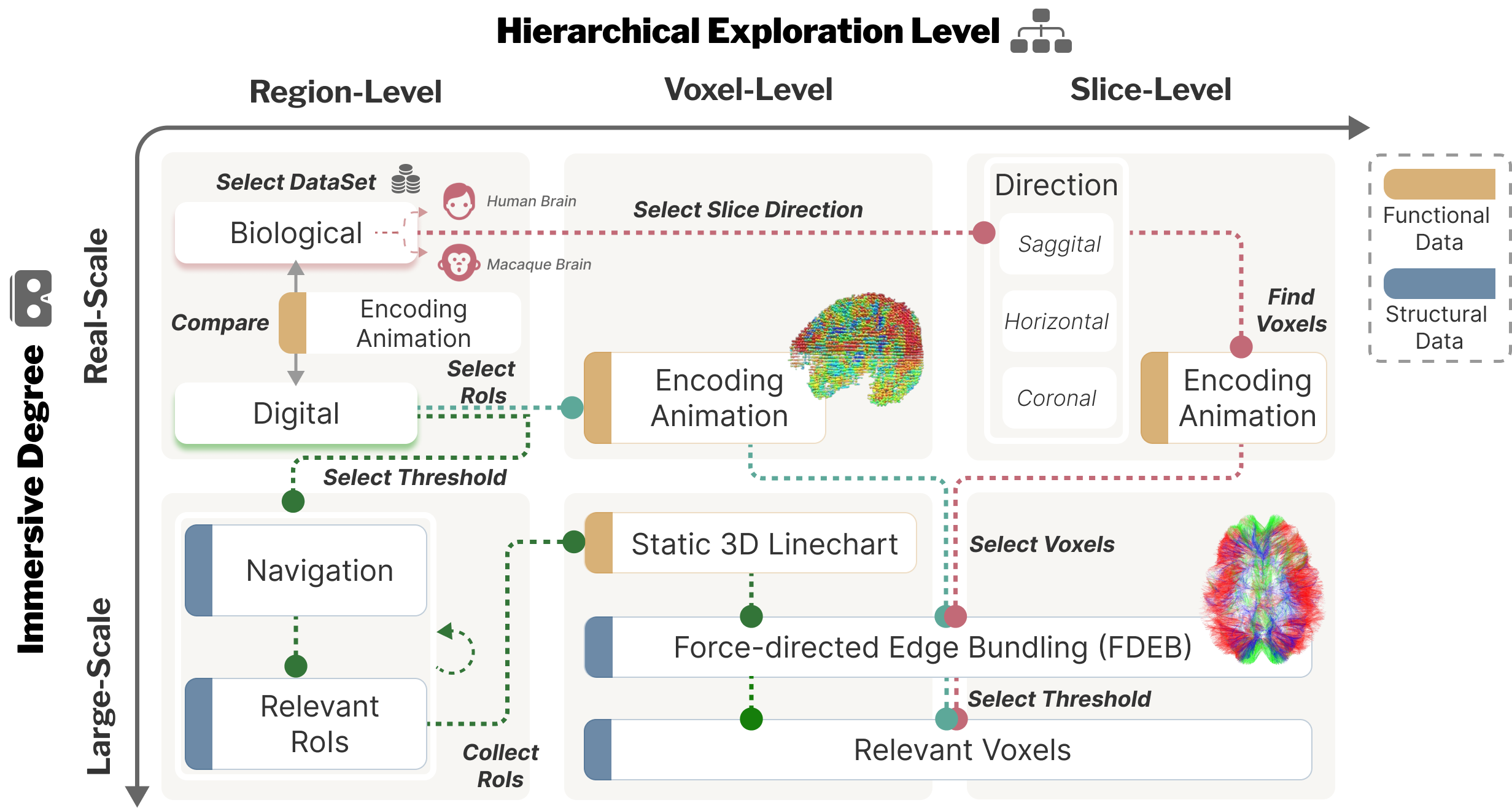}
    \caption{The figure illustrates the workflow of DTBIA, combining hierarchical exploration from region to voxel and slice levels with immersive navigation from real-scale to large-scale brain models. \textbf{The green exploration paths} begin with DTB data, comparing it to biological brain data to identify regions of interest (RoIs). Users can follow the deep green path to navigate the large-scale brain for relevant RoIs or take the light green path to move to the voxel level, where encoding animations and FDEB refine voxel selection. \textbf{The red exploration path} starts with biological brain data (human or macaque), leading to slice-level analysis. Through directional slicing, users examine voxel activity, then transition to large-scale brain models for further analysis using FDEB. Functional BOLD data is represented in brown, and structural DTI data in blue.}
     \label{fig:wf}
\end{figure*}

\subsubsection{Task Analysis}
Through extensive collaboration with domain experts, we identified key analytical tasks essential for Digital Twin Brain (DTB) research. These tasks address the challenges of processing, analyzing, and visualizing high-dimensional brain data to deepen our understanding of brain function and its underlying mechanisms. Based on expert feedback, we categorized them into the following areas:

\textbf{T1: Exploring Similarity Between Simulated and Real Neural Signals.} 
One of the primary tasks identified by the experts (E1, E2) is to assess how well brain-inspired models simulate real-world neural signals. E1 emphasized that, for complex data like DTB, visual comparison is more effective than traditional data analysis in assessing model fidelity. This method allows experts to intuitively identify differences and similarities between simulated and real signals, focusing on key metrics such as firing rates and time-series patterns.

\textbf{T2: Spatiotemporal Analysis of Functional Brain Data.} 
Experts (E1, E3, E5) emphasized the need for tools to identify regions of interest (RoIs) and analyze temporal patterns in brain activity, especially within functional data like BOLD signals. They require the ability to track how neural activity evolves over time, observe interactions between brain regions, and analyze correlations during cognitive tasks. These capabilities are essential for understanding brain function across different time scales and conditions.

\textbf{T3: Topological Analysis of Structural Brain Data.} 
Experts (E3, E5, E7) emphasized the need to identify the brain's topological structure and understand how regions are interconnected. The task involves analyzing connectivity patterns from methods like diffusion tensor imaging (DTI) to uncover the brain's network structure. Experts require tools that can display complex networks, identify key pathways, and explore the relationship between structural connectivity and brain function.

\subsubsection{Design Requirement}
Based on the task analysis, we derived the following design requirements to ensure that the system effectively supports the tasks identified by experts and addresses their specific needs:

\textbf{R1: Immersive Spatial Understanding of Brain Data (Supporting T1-T3).}
When analyzing complex neural data, such as spatiotemporal patterns\textbf{(T2)} or connectivity structures\textbf{(T3)}, it often involves tasks that exceed the capabilities of traditional 2D interfaces. While flat-screen setups allow users to rotate views and interact with brain structures using mouse interactions, they lack the intuitive spatial awareness~\cite{mania2010cognitive, rasheed2015immersive} required for effectively analyzing complex neural data. To address the inherent limitations of 2D displays, the system should provide advanced navigation~\cite{kim2024locomotion, rahimi2018scene} and interaction capabilities that facilitate seamless transitions between macro- and micro-level analyses. Systems that support six degrees of freedom (6DoF) for head movement offer a more immersive exploration, enabling users to navigate naturally within a 3D space~\cite{willemsen2009effects, sportillo2017immersive}. This approach facilitates a more intuitive understanding of brain structures, allowing for closer inspection of specific regions and a deeper comprehension of spatial relationships.

\textbf{R2: Hierarchical Exploration of Functional and Structural Data (Supporting T2-T3).}
Given the hierarchical nature of brain data and the need to manage its extensive scale, the system should offer a hierarchical approach to data exploration. For functional data\textbf{(T2)}, this means allowing users to move from broad overviews of brain activity to detailed, region-specific analysis, focusing on time-dependent patterns. For structural data\textbf{(T3)}, the system should enable exploration from high-level connections between brain regions down to individual Voxel-Level details. This hierarchical navigation ensures that experts can focus on key information at each level without being overwhelmed by the complexity of the entire dataset.

\textbf{R3: Managing Visual Complexity and Reducing Clutter (Supporting T3).}
The system should address the challenge of visual clutter, especially when analyzing large-scale, high-dimensional brain data. Dense neural connections often result in overlapping pathways, which can obscure important insights. The system should incorporate techniques that organize and simplify visualizations, such as consolidating overlapping edges to reduce noise. Additionally, interactive filtering tools should allow users to selectively explore specific regions or data types, ensuring that key patterns and connections are emphasized while minimizing distractions. This will help experts focus on the most relevant data during their analysis without becoming overwhelmed by the volume of information.

\section{Methodology and Immersive Design}
In this section, we introduce the workflow~\cref{fig:wf} and the immersive design of DTBIA, developed to explore BOLD signals and DTI data from the DTB. DTBIA enables researchers to explore brain region similarities, voxel-level connections, and structural pathways through immersive 3D visualization and hierarchical exploration.

\subsection{System Workflow}
The core design of DTBIA follows Shneiderman's "Overview first, zoom and filter, then details on demand" principle~\cite{shneiderman2003craft}, a widely recognized framework for complex data analysis, particularly suited for large datasets such as BOLD signals recorded across tens of thousands of voxels over multiple time points in brain research.

As illustrated in \cref{fig:wf}, the DTBIA workflow is structured along two key dimensions: hierarchical exploration process (left to right) and immersive interaction level (top to bottom). The hierarchical exploration process is divided into three phases — Region-Level, Voxel-Level, and Slice-Section Level — facilitating a smooth transition from broad overviews to detailed analyses, as depicted by the A-to-C exploration path in the teaser~\cref{fig:teaser}. 
On the immersive interaction level, DTBIA employs two distinct brain models: a Real-Scale brain for manipulation and interaction, and a Large-Scale brain for immersive navigation (represented on the far-screen side in the teaser~\cref{fig:teaser}(E)). This design enables users to seamlessly shift between distant observation and in-depth exploration, allowing for closer inspection by virtually navigating inside the brain model.

DTBIA integrates functional BOLD data with DTI-based structural connectivity for both DTB and biological brain datasets, enabling users to analyze neural activity alongside its underlying structural networks. At the Region-Level, using the Real-Scale brain model, users can compare simulated neural signals with real-world data \textbf{(T1)} to evaluate the accuracy of the DTB model and uncover additional patterns in both datasets. This phase also facilitates spatiotemporal analysis \textbf{(T2)}, where users can employ BOLD-encoded animations in the Real-Scale brain to identify regions of interest (RoIs). Upon identifying these regions, users can proceed to the Voxel-Level for more granular analysis or switch to the Large-Scale brain for immersive exploration, refining and collecting relevant RoIs iteratively. 

At the Voxel-Level, users can delve deeper into topological patterns \textbf{(T3)} by examining static 3D line charts of BOLD signals. This allows for the selection of key voxels and their associated connectivity, using the Force-Directed Edge Bundling (FDEB) algorithm with adjustable thresholding to isolate and focus on the most relevant voxel clusters. In the final Slice-Section phase, users can explore the biological brain slices through sagittal, horizontal, and coronal views, enabling precise examination of the brain's structure at the slice level. These views support a combination of voxel analysis via real-time encoded animations in the Real-Scale brain and the FDEB-filtered connectivity in the Large-Scale brain, enabling comprehensive spatial and functional investigations of neural data.

In the following subsection, we provide further details on the immersive interaction design, followed by an in-depth discussion of the hierarchical data analysis process in the next section.

\subsection{Exploration in an Immersive Environment}
Traditional desktop-based platforms~\cite{li2023dtbvis} often encounter significant occlusion challenges~\cite{elmqvist2008taxonomy, wang2019vr} when visualizing numerous voxels within a single brain region \textbf{(R3)}, leading to insufficient spatial awareness. To address this limitation, we propose an extension of our system into a virtual reality (VR) environment. This approach enables users to explore and interact with spatial voxel data from any position and angle, akin to real-world navigation, thereby offering a more immersive and engaging experience. Such a roaming capability enhances researchers' ability to uncover insights that may be difficult to obtain through conventional desktop platforms \textbf{(R1)}.

DTBIA leverages immersive virtual environments (IVEs) by adopting an egocentric perspective, embedding users directly within the data for intuitive exploration. To accommodate different spatial scales and exploration needs, DTBIA employs two brain models in its hierarchical workflow: a Real-Scale brain for an overarching view of brain data, and a Large-Scale brain model for detailed navigation and exploration.
The Real-Scale model supports egocentric walking, enabling users to physically navigate the brain's spatial representation, while the Large-Scale model enables egocentric flying, providing an unrestricted view for exploring large-scale structures. These approaches align with established immersive analytics (IA) navigation strategies, where both walking and flying allow users to manipulate and explore data effectively within IVEs\cite{wagner2021effect}. 
The dual-model strategy allows users to first acquire a comprehensive overview and then delve into specific areas of interest. Using animated navigation or manual exploration with VR controllers, users can "fly into" the brain and closely examine data points or voxels with fine granularity \textbf{(R2)}. This method addresses the common challenges of visualizing complex 3D structures on traditional 2D displays, particularly occlusion~\cite{wang2019occlusion, yilmaz2007conservative}, by promoting a more nuanced exploration of brain data\cite{ens2021grand}. Consequently, users gain a deeper understanding of the brain’s intricate spatial relationships.
By integrating these two modes, DTBIA enhances spatial awareness, improves task efficiency, and supports seamless navigation across multiple data scales.

\subsection{3D-FDEB for Structural DTI Data}
Building upon the immersive environment and hierarchical exploration process, DTBIA employs 3D Force-Directed Edge Bundling (3D-FDEB)~\cite{zielasko2016interactive, holten2009force} to reduce visual clutter in large-scale DTI data by clustering spatially and geometrically similar edges \textbf{(R3)}. The algorithm selects the top 10\% of connections based on significance (e.g., connection strength), extracts the x, y, z coordinates of the endpoints, and clusters the edges. The output is saved in JSON format, containing the coordinates of all points along each edge. This output enables immersive 3D visualization, where the points are connected to form continuous pathways. 

The force on a point \( F_{p_i} \) is given by:
\[
F_{p_i} = k_P \cdot (\|p_{i-1} - p_i\| + \|p_i - p_{i+1}\|) + \sum_{Q \in E} \frac{\|p_i - q_i\|}{C_e(P, Q)},
\]
where \( k_P \) controls the spring force, and the second term models electrostatic repulsion based on edge compatibility (\( C_e(P, Q) \)). This clustering process enhances the visualization of key structural connections, which are color-coded based on connection strength, making complex neural networks more interpretable and improving the clarity of white matter pathways.

\section{DTBIA System}
In this section, we detail the three hierarchical levels of exploration in the DTBIA system~\cref{fig:wf}. DTBIA allows users to explore brain data through three levels: Reion-Level, Voxel-Level, and Slice-Section-Level exploration. This hierarchical approach facilitates deeper analysis while maintaining clarity, helping users focus on key insights without being overwhelmed by the data's complexity. Additionally, DTBIA integrates cross-species brain models, including the macaque brain, enabling comparative research across species.

\begin{figure*}[htb]
     \includegraphics[width=\linewidth]{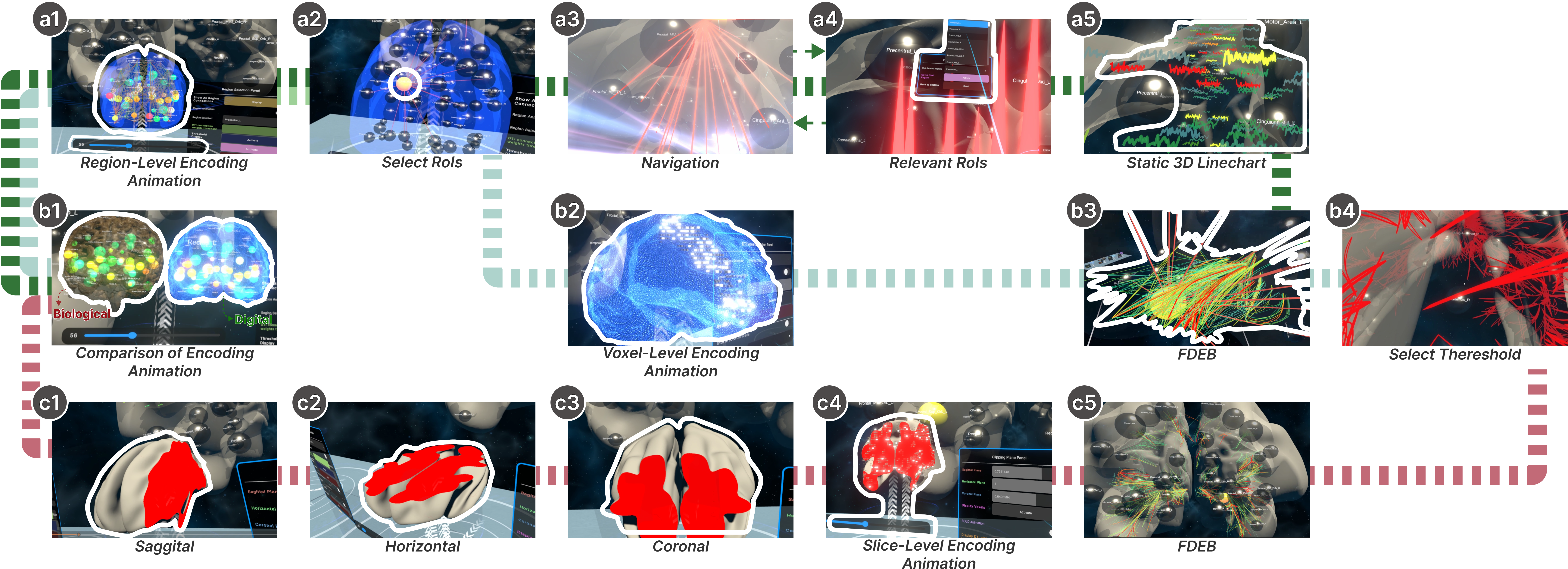}
    \caption{This figure illustrates three exploration paths of DTBIA. The green paths begin by comparing biological brain data with DTB data through encoding animations to assess model validity. The deep green path identifies peak brain activities with a time slider, selects RoIs, and navigates to the large-scale brain for further exploration, highlighting regions and connections. Users then analyze static 3D line charts and apply thresholding to the FDEB results, refining connections to show the most relevant links. The light green path diverges by focusing on Voxel-Level encoding animations, allowing for the identification of high-activity voxels and further refinement through FDEB and threshold selection. The red path, starting from biological brain data, involves Slice-Section-Level exploration, utilizing sagittal, horizontal, and coronal planes to find voxels within the slices, followed by encoding animations, FDEB, and threshold refinement.}
     \label{fig:pipe}
\end{figure*}

\subsection{Brain Region Exploration}
The exploration process begins at the Region-Level, where users interact with both a real-sized brain model and a large-scale navigation brain model to investigate brain regions in detail.

\subsubsection{Engagement with the Real-sized Brain Model.} The Region Exploration Level in DTBIA provides an interactive interface for users to investigate the brain's regional anatomy~\cref{fig:pipe}. Transparent spheres represent distinct brain regions, each labeled with its AAL (Automated Anatomical Labeling) tag. The size of each sphere corresponds to the voxel count of the respective region, visually indicating the region's volume. Below the brain model, a time slider allows users to visualize how BOLD signal activity fluctuates over time across these regions. As time progresses, the spheres change color from transparent to a full spectrum — green, yellow, and red — reflecting different levels of brain activity~\cref{fig:pipe}(a2). 
This interface also includes a side-by-side comparison mode\textbf{(T1)}, which displays biological brain data on the left and DTB data on the right~\cref{fig:pipe}(b1). 
Juxtaposition (i.e., side-by-side comparison)~\cite{gleicher2011visual} has been shown to be effective in providing clear and intuitive insights when comparing complex datasets, particularly in high-dimensional domains like brain activity signals.
By visually comparing both models, users can easily identify periods of peak activity and focus on regions that show consistently high BOLD signals, offering deeper insights into the brain’s functional behavior.

For the analysis of Diffusion Tensor Imaging (DTI) data, we have streamlined the process to emphasize the visualization of the most significant connections, highlighting the top 10\%. These connections are visually differentiated by a color gradient that transitions from green to orange, illustrating the directional flow between source and target regions. The panel on the right side incorporates interactive tools for an enhanced user experience, including a drop down list for direct region selection. Selecting a region triggers an update that highlights the chosen sphere in yellow and its connections in red~\cref{fig:pipe}(a3), observable in both the real-scale brain model and the large-scale brain. This immersive interaction facilitates a holistic view of the connectivity between different brain regions, further enriched by a threshold slider that filters connections based on their weight to underscore those of paramount importance.

\subsubsection{Navigation within the Large-Scale Brain.} After exploration in the real-scale brain model, users are encouraged to employ the teleportation feature to transition to their chosen region's site within the broader large-scale brain. This shift provides a more immersive, first-person perspective for exploring the network of connections emanating from the selected region. The Flying Navigation Panel appears upon reaching the region's sphere which introduces three pivotal tools: an auto-updating drop down for refined region navigation, an activation button for seamless transition to subsequent regions of interest, and a reset button to return to the exploration's starting point~\cref{fig:pipe}(a4). The drop down menu refreshes with each selection, ranking connections by their weight to guide the user towards areas of significant interconnectivity. This navigational feature offers an enriched understanding of the spatial arrangement and directional orientation of neural pathways \textbf{(T3)}, with areas previously explored being marked in yellow for quick identification. Upon completing their journey, users can utilize the reset option to revisit the starting point, where all explored regions are highlighted, facilitating a structured approach to further detailed investigation.

\subsection{Voxel-Level Exploration}
Building on the insights gained from the Region-Level exploration, the Voxel-Level analysis provides a more granular examination of neural activities and structural connections at a finer scale. At this level, DTBIA facilitates an in-depth investigation of both functional BOLD and structural DTI data:

\subsubsection{Exploring Functional BOLD Data.} Building on the regions of interest identified at the previous exploration level, we delve into the voxel exploration level, where DTBIA facilitates a detailed investigation. Within this framework, voxels within the selected region are represented as cubes, each symbolizing their biological significance. Users can adjust the time slider beneath the real-size brain model to dynamically visualize BOLD signal intensities within these voxels, employing the color mapping spectrum — from transparent black to emissive white — to indicate activity levels~\cref{fig:pipe}(b2). For more nuanced exploration, static line charts are positioned at each voxel's location within the navigation brain model. The color of each line chart transitions from green to yellow to red, reflecting the average BOLD signal across all time points~\cref{fig:pipe}(a5). This allows users to roam within the navigation brain model, identifying voxels of high activity by emmisve color and examining their time-series data for activity trends \textbf{(T2)}.

\subsubsection{Exploring Structural DTI Data.} When analyzing structural DTI data, we highlight the most significant connections, visualizing the top 10\%, or 38,036 links, to mitigate visual clutter from direct line connections. To achieve a clearer, more intuitive visualization, we employ a three-dimensional edge bundling algorithm that leverages force-directed layout principles. This algorithm iteratively optimizes the arrangement of connection lines, enhancing the clarity of data presentation. The connections are then color-coded according to their weight values. To investigate the DTI data further, users can select any region of interest identified earlier or display connections above a certain threshold to reveal areas with a high concentration of connections. Selecting a specific region in the navigation brain model will highlight connections based on voxel positions~\cref{fig:pipe}(b3), allowing users to observe connection patterns, including directionality and clustering, post-bundling. Additionally, an exploratory mechanism visualizes all connections initially, with a threshold slider provided to refine visible links~\cref{fig:pipe}(b4). Through this process, users can roam and identify areas with dense connections, selecting regions for deeper exploration.

\subsection{Slice-Section Exploration}
Following voxel-level analysis, slice-section exploration allows for deeper anatomical insights by focusing on specific planes, revealing spatial distributions and connectivity patterns across different brain sections.

To gain deeper biological insights, the slice-section feature allows users to examine specific brain regions by slicing the brain at the region of interest. Building on the initial display of brain regions, users can select and display sagittal, horizontal, or coronal planes~\cref{fig:pipe}(c1-c3), depending on the position of the region they wish to explore. The sagittal plane divides the brain into left and right hemispheres, the horizontal plane into upper and lower parts, and the coronal plane into front and back sections. These slice views generate two-dimensional images of the brain at specific time points, revealing the spatial distribution of BOLD signals through a color map that transitions from transparent black to emissive white~\cref{fig:pipe}(c4). Adjusting the current time point allows for dynamic observation of changes in the color map, indicating fluctuations in BOLD signal intensity. Areas marked in red on the color map signify heightened neural activity, providing valuable insights into the spatial dynamics of BOLD signals within the brain.

The hierarchical exploration process — starting with broader brain regions and progressing to more detailed voxel and slice analyses — enables users to discern spatial patterns of brain activity across different planes. Comparative analysis of voxels in the sagittal and horizontal planes reveals anterior-posterior patterns, while the horizontal and coronal planes highlight lateral distribution, and the sagittal and coronal planes offer insights into vertical patterns. Additionally, slice-section exploration extends to the visualization of DTI connections originating from voxels within the sliced sections of the brain~\cref{fig:pipe}(c5). This feature allows users to examine the direction, density, and distribution of neural pathways, providing a comprehensive view of both the functional and structural connectivity of the brain. By integrating DTI visualization into the slice exploration process, DTBIA enhances the depth and scope of neuroscientific analysis, offering a more holistic understanding of brain connectivity and activity.

\begin{figure}[h]
    \centering
    \includegraphics[width=\columnwidth]{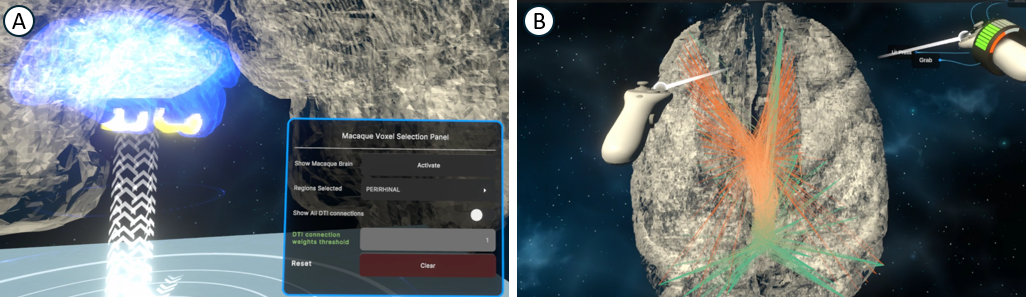}
    \caption{This figure demonstrates the system's capability to integrate macaque brain data, showing (A) functional data by region and (B) DTI connections among voxels within a region.}
    \label{fig:sys_fig4}
\end{figure}

\subsection{Integration of Macaque Brain Data}
DTBIA extends its capabilities beyond human brain data to include digital models of other species, showcasing the system's versatility and its potential to broaden the scope of neuroscience research. A prime example of this adaptability is the incorporation of macaque brain data, which enables comparative neuroanatomical studies and offers insights into the evolutionary aspects of brain development. Through interactive exploration features, researchers can analyze both functional data ~\cref{fig:sys_fig4}(A) and structural~\cref{fig:sys_fig4}(B) of the macaque brain, mirroring the exploration levels provided for human data. This inclusion not only facilitates a deeper understanding of primate neurobiology but also enhances cross-species research, allowing scientists to uncover universal principles of brain structure and function. As DTBIA continues to evolve, the integration of additional species' brain data promises to enrich comparative neuroscience studies further, making it an invaluable resource for both human and animal brain research.
\section{Case Study}
This section presents detailed descriptions of the two case studies: the first without biological domain expertise to assess core functionality, and the second with biological experts to explore deeper, reflecting the interdisciplinary nature of brain-inspired research. To ensure an objective assessment, the case studies involved both internal and external experts. In total, 12 experts participated in the evaluation: two from the original group (E1 and E5) and two external experts (E11 and E12). Experts E1 and E11, specializing in brain-inspired computing, contributed to the first case study, while E5 and E12, with expertise in biological neuroscience, were involved in the second case study. The remaining eight experts from the original group provided additional feedback by observing the VR interactions of the selected experts (E1 and E5) via screen streaming. These two case studies were conducted on a desktop system equipped with an Intel i5-12490F CPU and an NVIDIA GeForce RTX 3060 Ti GPU. The system used SteamVR for streaming and a Quest 3 headset as the head-mounted display (HMD).

\subsection{Exploration without biological domain knowledge}
In the first case study, we assisted the brain-inspired computing experts (E1, E11) in exploring functional magnetic resonance imaging (fMRI) and structural DTI data using the DTBIA system. The experts' primary objective was to assess the validity of the simulated results and identify regions and voxels of high neural activity. The exploration began at the Region Exploration Level, where the experts utilized the animation playback feature to visualize the fMRI data. In this process, the experts compared the BOLD signals from the biological brain with those from the DTB. During the side-by-side comparison of BOLD signal intensities, a slight delay in the response of the DTB was observed. Specifically, peak activity in the DTB occurred approximately three time points later than in the biological brain. Furthermore, the signal intensity in the DTB was weaker, as indicated by the color encoding scheme used in the visualization. The experts identified that this discrepancy may be due to the limitations of the DTB model, which likely does not fully replicate the complex electrochemical processes of the biological brain or account for all biological details accurately.

\begin{figure}[h]
    \centering
    \includegraphics[width=\columnwidth]{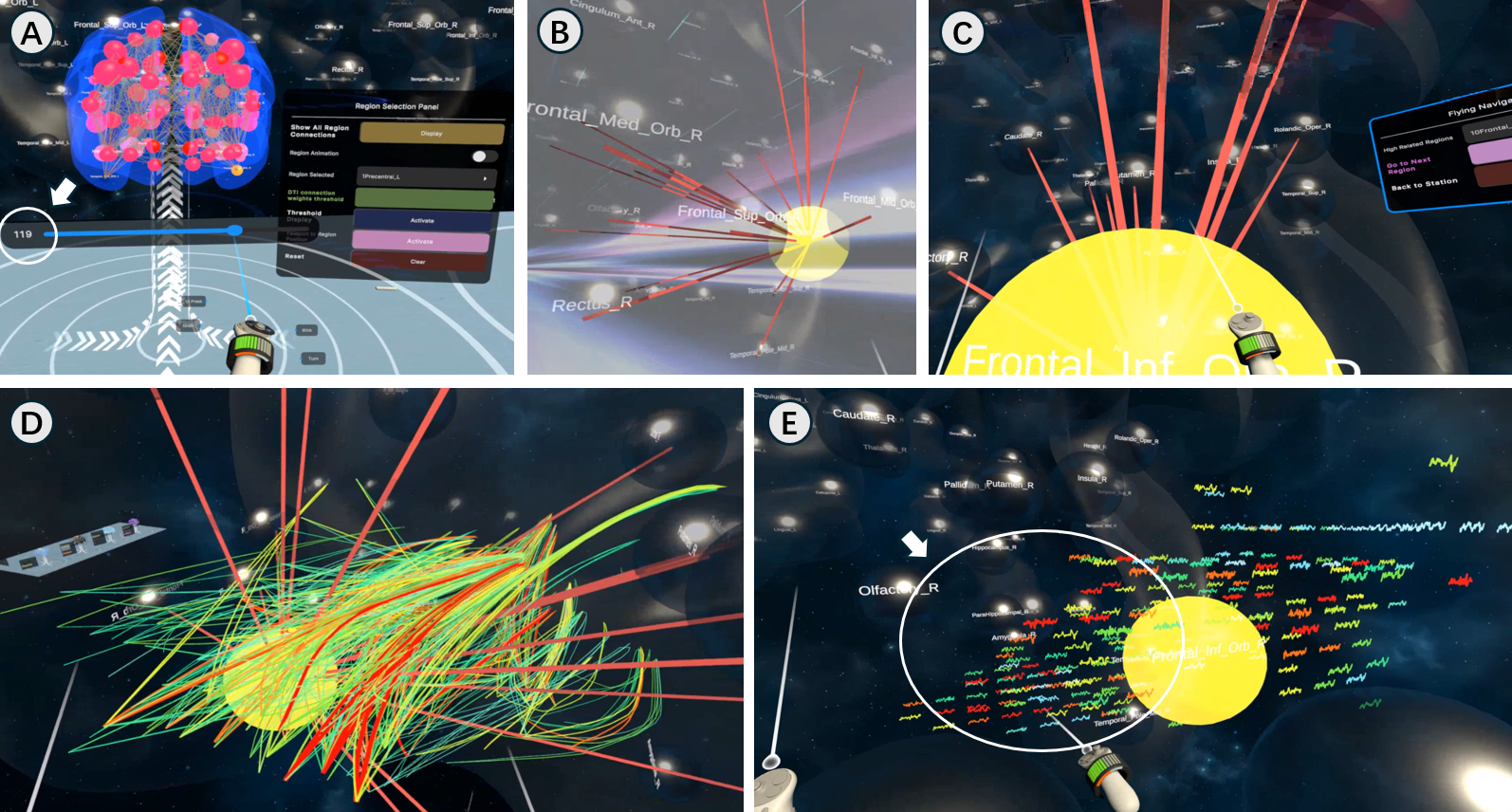}
    \caption{This figure showcases key findings from the first case study: (A) Time slider use reveals peak brain activity at time point 119. (B) Region 16 (Front\_Inf\_Orb\_R) shows high BOLD signals, leading to a teleportation to it. (C) Explores extensive connections in this region. (D) Reveals DTI connections among voxels. (E) Line charts near the brain's center highlight above-average activity.}
    \label{fig:fig6}
\end{figure}

To identify regions of high neural activity, the experts began by examining the overall brain activity. At time point 119 \cref{fig:fig6}(A), peak activity was observed, indicated by the majority of spheres turning red. To refine the analysis further, the experts adjusted the connection weight threshold on the right panel to 0.8, isolating the top 20\% of connections and focusing on the most dynamically connected regions of the brain. Through this focused analysis, the Right Inferior Frontal Gyrus, Orbital Part (Front\_Inf\_Orb\_R), located in the lower frontal region of the right hemisphere, was identified. The experts then teleported to this region within the Large-Scale navigation space~\cref{fig:fig6}(B), immersively exploring its connections and noting that they predominantly extended toward the occipital area~\cref{fig:fig6}(C). To further investigate, the experts advanced to the Voxel Exploration Level, initially exploring DTI connections among voxels and navigating through them using the VR controller's thumbsticks. After examining the connections from multiple angles, the experts observed that most of them had a lateral (left-right) fiber orientation~\cref{fig:fig6}(D). Additionally, by analyzing the line charts corresponding to the voxels — which displayed the average BOLD signal intensity through color gradients — the experts noted that the majority of line charts clustered near the brain's center, indicating areas of above-average activity~\cref{fig:fig6}(E).

Our proposed system offers a user-friendly and intuitive approach for domain experts to gain valuable insights from fMRI data, enabling them to efficiently identify regions of interest (ROIs) and focus on specific voxels for further analysis. Based on the feedback received from the experts, the experts initially expressed the positive evaluation of the system's interface representation, noting its overall conciseness, clarity, and user-friendliness. Moreover, they commended the innovative exploration capabilities of the VR environment \textbf{(R1)}, as it provided users with greater flexibility to move around and observe spatial information of interest at a closer range through roaming and head movement. Finally, they acknowledged that the overall exploration process of the system facilitated expert identification and extraction of critical regions and voxel points for subsequent analysis. Overall, their feedback indicates that our system has significant potential to enhance the efficiency and effectiveness of DTB analysis within a VR environment.

\subsection{Exploration with biological domain knowledge}
In the second case study, we collaborated with the neuroscientists (E5, E12) possessing a solid background in biology and prior expertise in the functions of individual brain regions. The experts' goal was to explore the whole-brain resting-state BOLD signal fMRI data, with a particular focus on understanding the Default Mode Network (DMN). The DMN is a network of brain regions that becomes active when an individual is not focused on the external environment, but rather on introspection, daydreaming, or mind-wandering. It typically includes the following seven regions, with their corresponding region labels in parentheses: 
Frontal\_Sup\_Medial(23, 24), Hippocampus(35, 36), Amygdala(39, 40), Parietal\_Inf(59, 60), Precuneus(65, 66), Thalamus(75, 76), Temporal\_Pole\_Sup(81, 82).
These regions work in coordination to support a range of cognitive processes, including self-referential processing, social cognition, and memory retrieval, among others. 

\begin{figure}[h]
    \centering
    \includegraphics[width=\columnwidth]{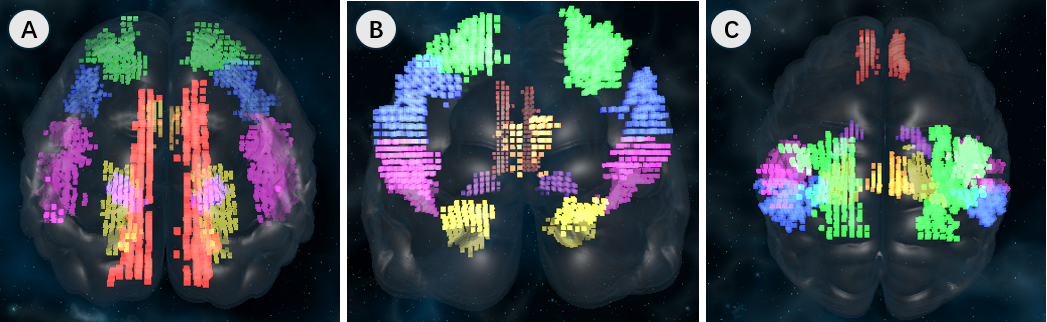}
    \caption{We use different colors to display each of these 7 regions in a graphical representation of the DMN to make it easier to distinguish and analyze them. Results can be viewed from different angles: (A) Front view (B) Back view (C) Top view }
    \label{fig:case2_1}
\end{figure}

To initiate the exploration process, the experts first annotated the locations of these 14 brain regions. By using different colors for each region, the relative positions of different brain areas could be seen in ~\cref{fig:case2_1}. During the Region Exploration phase, the experts used the size of the spheres to assess the number of voxels in each region, while the color intensity indicated the average BOLD signal for each region. Analysis revealed that the temporal lobe had the highest number of voxels, with 626 across both hemispheres, followed by the parietal lobe with 596 voxels. These regions are significantly larger compared to others. In terms of neural activity, the hippocampus exhibited the highest average BOLD signal (0.02433), closely followed by the temporal lobe (0.0243). Both regions had higher average BOLD values than the overall brain average, indicating their active involvement during the resting state.

\begin{figure}[h]
    \centering
    \includegraphics[width=\columnwidth]{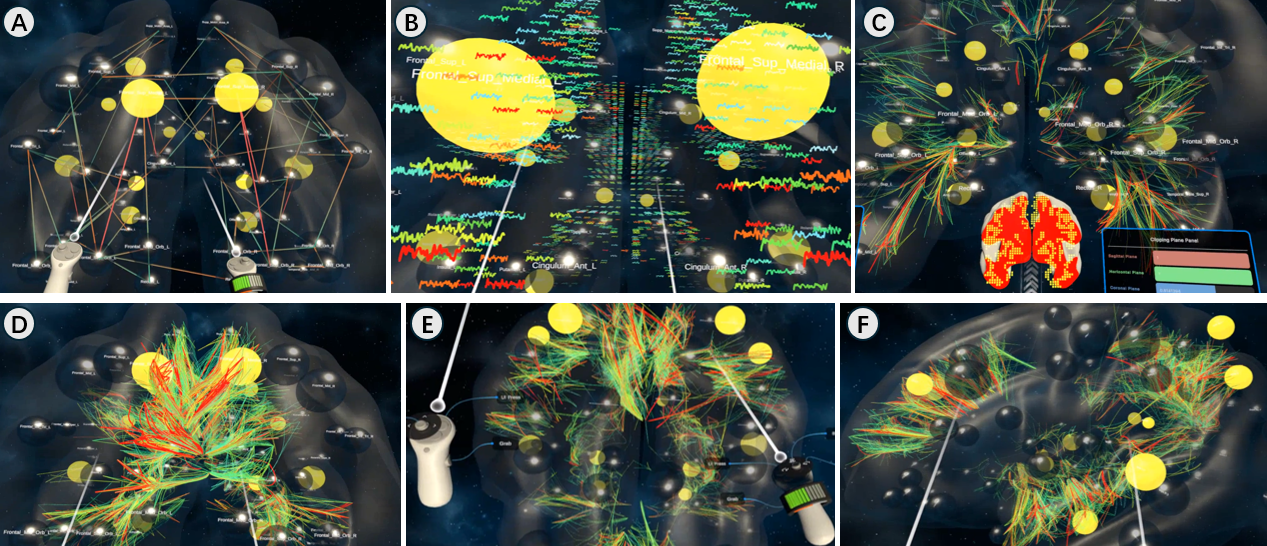}
    \caption{This figure showcases the second case study's findings in the DMN: (A) Identifies DMN regions. (B) Displays voxel line charts. (C) Uses slicing sections for anatomical analysis. (D-F) Show DMN from front, top, and back views, revealing its structure and connections. }
    \label{fig:case2_2}
\end{figure}

To further investigate the DMN, the experts selected and highlighted all these region spheres to start the navigation~\cref{fig:case2_2}(A). The experts also visualized the line charts for each voxel's BOLD signal beneath these regions~\cref{fig:case2_2}(B), and found that the Hippocampus had the highest average BOLD voxels. In the biological explanation, the Hippocampus was found to be involved in DMN activity during memory retrieval and imagination. The experts then examined the DMN's structure by visualizing the DTI connections from these regions' voxels. This analysis was conducted from three different angles~\cref{fig:case2_2}(D-F), providing a comprehensive view of the connections. Given the hippocampus' functional importance in the DMN, the experts used the Slice Section view~\cref{fig:case2_2}(C) to better understand its anatomical context. By leveraging the detailed voxel information provided in the Slice Section view, the experts identified key voxels for in-depth analysis within their field of expertise.

Based on expert feedback, the DTBIA system proved effective for exploring both functional and structural aspects of brain data, particularly the DMN. Experts used the system to investigate neural activity in key DMN regions, gaining insights into their role in cognitive processes like introspection and memory. The experts noted that the system provided an intuitive, interactive platform for exploring complex brain data, enhancing understanding of the spatial distribution and connectivity of the DMN. Additionally, it allowed experts to validate their biological knowledge by comparing real and simulated data, improving the accuracy of their conclusions. The experts also noted that this iterative, data-driven exploration helped bridge the gap between their traditional, more static analysis methods and the dynamic, immersive interaction provided by the DTBIA system. By facilitating better collaboration with brain-inspired computing experts, the system allowed for a more integrated and multi-faceted approach to understanding brain function, which had previously been more one-sided.

Through these two case studies, we have demonstrated the effectiveness of the DTBIA system in exploring brain data. Experts' feedback highlights the system’s capacity to deepen understanding of neural activity and facilitate cross-disciplinary collaboration. In the following section, we summarize these insights and outline potential directions for future development.

\section{Discussion And Future Work}
In this section, we discuss the rationale, usability, and significance of DTBIA, assess its limitations, and propose directions for future research to enhance its capabilities and impact.

\subsection{Discussion}
This paper introduces DTBIA, an immersive visual analytics system designed to support brain-inspired research by facilitating interactive exploration of the Digital Twin Brain (DTB). The system integrates large-scale brain data, enabling the analysis of spatiotemporal patterns, visualization of brain structure and function, and investigation of brain activity at various levels of granularity. Tailored for domain experts such as neuroscientists, biologists, and AI researchers, DTBIA equips users with tools to explore brain dynamics, refine computational models, and develop hypotheses based on emerging insights. Its domain-specific design optimizes the system for investigating neural mechanisms, cognitive processes, and brain networks, positioning it as a valuable resource for advancing brain-inspired research.

Experts' feedback indicates that DTBIA's immersive capabilities are particularly beneficial in enhancing communication between researchers from diverse disciplines. The ability to visualize complex brain data in an interactive 3D space allows neuroscientists to better understand and interpret simulation results, particularly those related to brain activity and signal transmission. By presenting these findings in an intuitive, visual format, DTBIA facilitates clearer communication of complex concepts, fostering collaboration and idea exchange between biologists, neuroscientists, and AI experts.

DTBIA's cross-platform compatibility further expands its applicability across a variety of research and practical contexts. The system can be deployed on both Windows and Android platforms, making it suitable for use in academic laboratories, clinical environments, and outreach programs. This flexibility enables DTBIA to engage both academic researchers and the general public, promoting an immersive and accessible way to explore brain function. By facilitating cross-disciplinary collaboration between neuroscience, biology, and AI, DTBIA supports the development of integrative models of brain function, advancing research in both fields. As the system evolves, it will continue to drive innovations in brain science and AI, particularly in areas such as brain disorders, cognitive functions, and brain-computer interfaces.

\subsection{Limitation And Future Work} 
While DTBIA has proven valuable in analyzing spatiotemporal patterns and advancing brain function understanding, several limitations remain. This section addresses challenges related to system performance, user experience, and data scope, which are essential for gaining deeper insights.

\textbf{System Performance and User Experience.}
One of the primary challenges lies in rendering performance. While the system typically runs at 90 FPS on an RTX 3060 Ti GPU, setting a low threshold leads to a significant increase in the number of objects (points and lines) being rendered in the scene. This overload can cause the FPS to drop below 10, potentially leading to cybersickness, especially in immersive environments. With an optimized threshold, when the total number of objects is kept below 10,000, the average FPS stabilizes around 45, which is sufficient for rendering DTI connectivity and BOLD animations without noticeable disruption.
Cybersickness is another concern during DTBIA’s exploration process due to the significant movement involved. Although mechanisms such as reducing the field of view (FOV) during movement have been implemented to minimize discomfort, further improvements are needed. For instance, smoother transitions, reduced motion intensity, and user-adjustable navigation options could mitigate cybersickness, ensuring a more comfortable and accessible experience for users.
Another limitation is the potential for subjective biases introduced by the system’s interactive exploration method. While the interactive nature of DTBIA allows for intuitive, in-depth analysis, it also depends on the user's choices and interpretations, which can lead to biases. For example, users might prioritize certain data points or regions based on personal interest or prior knowledge, potentially influencing the results. To mitigate this, further research is needed to assess the reliability and objectivity of the findings generated by DTBIA.

To enhance DTBIA’s potential, future work will focus on incorporating multiplayer functionality for real-time, embodied collaboration among cross-disciplinary experts. Currently, the system lacks this feature, limiting its ability to facilitate interactive, real-time engagement between biological and AI researchers. By enabling multiplayer interaction, we aim to create a more integrated platform where experts can collaborate in real-time, exchange insights, and engage in interactive analysis, fostering progress at the intersection of biological and artificial intelligence.
Additionally, the system's interaction capabilities can be improved to reduce the learning curve for domain experts. While scene resizing is currently achieved through camera translation, and gesture-based interactions are not yet implemented, these methods may not provide the most intuitive or natural experience. Future work should incorporate gesture controls and VR controllers to enable more natural, embodied interactions, making the system more accessible and user-friendly, particularly for experts with limited VR experience.

\textbf{Data Scope and Analytical Depth.}
Beyond system performance, the scope of the data utilized in DTBIA presents additional limitations. Currently, the system relies on 166 time-series data points reflecting the brain's resting state, as provided by the domain experts. While this dataset provides valuable insights into the brain's activity, expanding it to include additional brain states, such as those related to activities like eating or reading, could enhance our understanding of brain function. Furthermore, capturing neuronal-level dynamics, which are critical for deeper insights into brain function, remains a challenge. The current data, spanning from brain regions to voxels and slice sections, is insufficient to fully represent the brain's network at the neuronal level. As recent studies~\cite{potjans2014cell, schmidt2018multi} have highlighted, understanding brain activity at the neuronal level is crucial for advancing brain-inspired research and developing accurate models of neural processing. Future work should focus on enhancing the cortical network model by incorporating neuron-specific data, enabling more detailed exploration of brain activity and providing a foundation for modeling brain processing at the cellular level. Ultimately, these efforts will contribute to advancing brain research and fostering innovative breakthroughs in the field.

Overall, DTBIA provides a robust platform for investigating brain connectivity and function, with important implications for neuroscience. Future research should build upon these findings and focus on addressing the system's limitations to further strengthen its role in advancing both brain-inspired research and AI applications.
\section{Conclusion}
We have designed and implemented DTBIA, an immersive interactive visual exploration system for DTB research. The system was introduced and tested using resting-state BOLD signal and DTI data generated by DTB from the real world. Feedback from experts confirmed our hypothesis that integrating VR technology enables deeper and more unrestricted exploration of data behavior, offering a novel, experiential perspective in an immersive environment. This approach allowed for the analysis of spatiotemporal data patterns with a physical sense of space. Experts found the system to be a valuable asset for their models. Through our case study and expert validation, we have provided strong evidence supporting the practicality and effectiveness of our system.

\acknowledgments{
    This work is supported by Natural Science Foundation of China (NSFC No.62472099, No. 62302422 and No.62202105). This work is also partially supported by CORE, which is a joint research center for ocean research between Laoshan Laboratory and The Hong Kong University of Science and Technology.
}

\bibliographystyle{unsrt}
\bibliography{main}

\end{document}